# Mean square winding angle of Brownian motion around an impenetrable cylinder

J.H. Hannay and Michael Wilkinson*

H.H. Wills Physics Laboratory, University of Bristol, Tyndall Ave, Bristol BS6 7LP, UK.
*Department of Applied Mathematics, Open University, Walton Hall, Milton Keynes MK7 6AA, UK.

*Abstract*
An exact formula is derived, as an integral, for the mean square value of the winding angle $\phi$ (with $-\infty < \phi < \infty$) of Brownian motion (that is, diffusion) after time $t$, around an infinitely long impenetrable cylinder of radius $a$, having started at radius $R(>a)$ from the axis. Strikingly, for the simpler problem with $a=0$, the mean square winding angle around a straight line, is long known to be instantly infinite however far away the starting point lies. The fractally small, fast, random walk steps of mathematical Brownian motion allow unbounded windings around the zero thickness of the straight line. A remedy, if it is required, is to accord the line non-zero thickness, an impenetrable cylinder, as analysed here. The problem straightaway reduces to a 2D one of winding around a disc in a plane since the axial component of the 3D Brownian motion is independent of the others. After deriving the exact mean square winding angle, the integral is evaluated in the limit of a narrow cylinder $a^2 << R^2$, highlighting the limits of short and long diffusion times addressed by previous approximate treatments.

*Introduction*

A freely diffusing point in 3 dimensions, that is, one executing mathematical Brownian motion has mean square displacement Cartesian components $\langle\Delta x^2\rangle = \langle\Delta y^2\rangle = \langle\Delta z^2\rangle = 2D\Delta t$ in an infinitesimal time $\Delta t$, where $D$ is the diffusion constant, leading to an isotropically spreading Gaussian probability density. The projection of the motion onto, say, the *x,y* plane, is thus planar diffusion with the same diffusion coefficient. In polar coordinates, cylindrical or planar, the angle coordinate diffusion depends on the radius coordinate *r* through

$$\langle\Delta\phi^2\rangle = 2D\Delta t / r^2 \qquad (1)$$

and *r* is itself changing, so the angular spreading is more complicated than that of the Cartesian components. This polar angle can be of interest if the central axis (around which the angle is measured) has topological significance, as in the examples mentioned below. For this, one allows the polar angle, the winding angle of the trajectory about the axis (or the point origin in 2D), the freedom to vary from $-\infty$ to $\infty$ rather than being restricted to a range of $2\pi$. Optionally one can visualize the 2D diffusion as taking place on Riemann's flattened helicoid surface, free to wind to its different sheets.

It would be expected that this problem of polar coordinate evolution, being so easily stated mathematically, might have physical interest. Such winding problems do indeed arise: in optics (edge diffraction), quantum mechanics (the Aharonov-Bohm effect), and polymer physics (linking entanglement). Historically there were several largely independent (though closely related) mathematical analyses [Spitzer 1958](diffusion),[Aharonov and Bohm 1959](quantum mechanics), [Ito and McKean 1965](diffusion), [Edwards 1967](polymers) in these fields around the middle of the last century, all in apparent ignorance of the pioneering 'complex plane of unrestricted angle' method developed for optics by Sommerfeld [Sommerfeld 1896, 1964]. In all these, the need for polar coordinates arises because there is a point-like (in 2D) or straight-line-like (in 3D) feature or obstacle around which the winding is of interest. (The 2D and 3D versions are either equivalent or very closely related in all these different contexts). Another slightly more recent point-winding work, notable for its unusual point of view is that by Michael Berry [Berry 1980] based on a favourite technique of his: using the Poisson summation formula to convert between an angular momentum sum and a topological windings sum. Recent generalizations of point or line-like winding problems with exact results are to be found in [Hannay 2019][Hannay 2019].

There is a long known but particularly striking feature of polar angle evolution in this free Brownian motion (that is without obstructions); its mean square angle change after any time $t>0$, however small, is infinite. This arises from the fractal nature, the infinitesimal, infinitely frequent, random walk steps of Brownian motion, and applies whatever the starting radius $R$ is from the origin (or axis). Formally the infinity comes about because the spreading Gaussian probability density centred at a point at radius $R$ is non-zero at the origin for any $t>0$. This suffices: the rate of increase of the mean square angle $\langle \phi^2 \rangle$ is, from (1) together with the fact that $\langle \phi \Delta \phi \rangle = 0$, the integral over all $r > 0$ of $2D/r^2 \times 2\pi r dr \times$ probability density, which therefore diverges non-integrably at the origin.

This possibly unphysical feature is remedied, in perhaps the simplest way, by according a non-zero radius $a$ to the obstacle (point or line), creating an impenetrable reflective cylinder (or disc in 2D). Mathematically this means that the probability density obeys the Neumann condition of zero gradient, zero flux through the boundary. The corresponding problem with Dirichlet boundary condition of perfect absorption, which therefore loses probability, was studied somewhat differently [Rudnick and Hu1987]. Some subsequent literature, for both Neumann and Dirichlet boundary conditions, contains assorted approximate treatments, including long and short diffusion times [Grosberg and Frisch 2003], [Smith and Meerson 2019], [Huber and Wilkinson 2019]. The last of these, which treats the Neumann case, provides precise agreement with the long time limit (13) of our exact result (8), and is rederived in appendix A under the title 'truncated free space model'.

In the present paper we supply an exact formula, as an integral, for the mean square winding angle, starting at a point at radius $R$, with the Neumann boundary condition applied on the disc (or cylinder) boundary with radius $a$ centred on the origin. In the limit of a narrow cylinder $a^2 \ll R^2$ the exact formula integral can be evaluated in terms of special functions, and simplified further in the two limits of short and long diffusion times. For the narrow cylinder, the exact result is matched pretty well by the 'truncated free space model' (specifically, by equation (A2) in appendix A).

Brownian motion, that is, diffusion, has, via the formula (1) for the mean square change in winding angle, a uniquely simple consequence for the accumulated total mean square winding angle $\langle \phi^2 \rangle$. It is not shared by other statistics, for example $\langle \phi^4 \rangle$. The Markov property of the motion (independence of the future on the past) makes the accumulation simply algebraically additive $\langle (\phi + \Delta\phi)^2 \rangle = \langle \phi^2 \rangle + \langle \Delta\phi^2 \rangle$. Suppose that after time $t$ the diffused probability density $P(\mathbf{r},t)$ is known, and the $\langle \phi^2 \rangle$ is also known. Then the mean square angle $\langle (\phi + \Delta\phi)^2 \rangle$ after time $t+\Delta t$ instead of $t$, is dictated by $P(\mathbf{r},t)$. It supplies, via (1), the increment of $\langle \Delta\phi^2 \rangle$, re-starting from each point $\mathbf{r}$ reached after time $t$, weighted by $P(\mathbf{r},t)$.

As a very brief example, one can straightaway extract the leading term in the long time limit $t \to \infty$ since eventually the $P(\mathbf{r},t)$ approximates a Gaussian $\exp[-r^2/4Dt]/(4\pi Dt)$ that has spread to be so wide that its centre can be taken as the disc centre. Then from (1),
$$d\langle \phi^2 \rangle / dt \approx \int_a^\infty \exp[-r^2/4Dt]/(4\pi Dt)(2D/r^2)2\pi r dr$$
$$= E_1(a^2/4Dt)/(2t) \approx -\log(a^2/4Dt)/(2t) \text{ for large } t.$$ The indefinite integral of this over $t$ is $\log^2(a^2/4Dt)/4 + constant$. With a suitable choice of constant this matches the leading term from the exact analysis (13) from (8). The growth of the mean square winding angle is thus very slow in comparison with the linear growth of mean square displacement.

So the central ingredient for finding the mean square winding angle will be the probability density $P(\mathbf{r},t)$ in the plane, spreading by diffusion, having started at $t=0$ as delta function at a point at radius $R$. This diffusion takes place in the presence of an impenetrable disc of radius $a$ with the consequent Neumann boundary condition of zero radial derivative of $P(\mathbf{r},t)$ there. Actually, the circular symmetry of the disc means that the considerably easier probability distribution $\overline{P}(\mathbf{r},t)$ from an initial ring $\delta(r-R)/2\pi R$, of radius $R$ suffices in place of $P(\mathbf{r},t)$. This azimuthal average $\overline{P}$, the zeroth component of the angular decomposition of $P$, is the only feature required, because $P$ is to be weighted and integrated (2) with the isotropic function $1/r^2$ arising from the angular diffusion (1). The initial ring spreads (both inward and outward) by diffusion, producing the circularly symmetric probability density $\overline{P}(\mathbf{r},t)$, given in (6) below and accessed as described in the next paragraph. Finally, once the weighted integral of $\overline{P}(\mathbf{r},t)$ is found, a time integral of the rate of increase of the mean square winding angle over the duration of the Brownian motion will yield the desired mean square winding angle (2).

The circularly symmetric spreading probability density $\overline{P}(\mathbf{r},t)$ is to be accessed via its Fourier transform $\Psi(\mathbf{r},\omega)$ with respect to time, which is also circularly symmetric (the specific relationship is (4) below). This Fourier transform, incorporating a zero probability density for negative time, is complex like a wave. In fact it has a direct physical interpretation in terms of a more familiar problem of wave scattering by an obstacle. The wave here is the one that comes from a ring source of radius $R$ with a strength that continuously oscillates at a definite angular frequency $\omega$ (like a vertically vibrating circular dipper in a shallow water wave 'ripple tank'). Wave scattering theory prescribes the desired solution straightforwardly in principle, though with awkwardly long intermediate formulas like (5) below. An illustrative example of the wave method in a simpler context (no disc) is provided in appendix B. An alternative, scattering-free, route to the same result for $\overline{P}$ is derived in an appendix C, taking the limit of a large finite domain (an annulus).

The wave $\Psi(\mathbf{r},\omega)$ obeys the equation $\nabla^2\Psi + k^2\Psi = -\delta(r-R)/(2\pi R)$ where the right hand side represents a ring source of waves. This equation is the time Fourier transform of the diffusion equation with a δ-ring of probability introduced at $t=0$: $D\nabla^2\overline{P} = \partial\overline{P}/\partial t - \delta(t)\delta(r-R)/2\pi R$. The $k^2$ term on the left comes from the Fourier transform of ($\partial/\partial t$) with a change of variables $i\omega=Dk^2$. (Although physically $k$ can be interpreted as a wavenumber, with $i\omega=Dk^2$ as the diffusion dispersion relation, it is mathematically just change of variable; the Fourier transform is in time, not space). The boundary condition on the disc is the same one for both problems, the zero normal derivative, Neumann condition. There is an additional condition on $\Psi(r,\omega)$ arising from the vanishing of probability for negative time. It has already been incorporated into the description in terms of a 'source' rather than a 'sink' of waves. Since the waves come from a source (the oscillating ring) they must obviously be outward-moving outside $R$ (that is, $\propto H_0^{(1)}(kr)\exp(-i\omega t) \sim (1/\sqrt{kr})\exp(ikr-i\omega t)$ for large $r$). Inward moving waves would correspond to zero probability density for positive time instead.

*Exact formula*

For Brownian motion in 2D, the diffusing probability density $P(\mathbf{r},t)$, its mean square winding angle in the presence of an impenetrable disc of radius $a$, for a motion of duration $t$ starting at radius $R$ is given, following (1), by

$$\left\langle \phi^2 \right\rangle = \int_0^t dt' 2D \int_{|\mathbf{r}|>a} P(\mathbf{r},t') \frac{d^2\mathbf{r}}{r^2} \tag{2}$$

This representation was that used in a recent approximate 'truncated free space' model [Huber and Wilkinson 2019] reproduced in appendix A. It is specific to the mean square statistic (as mentioned above) and seems more manageable, analytically, than the more flexible one used by [Rudnick 1987] for the same problem with an absorbing boundary condition. The latter involves awkward derivatives of Bessel functions with respect to their order.

As explained in the introduction the probability density can be replaced without approximation by its azimuthal average, which is that from an initial ring instead of the initial point

$$\overline{P}(\mathbf{r},t) \equiv \frac{1}{2\pi r} \int_{|\mathbf{r}'|>a} P(\mathbf{r}',t) \, \delta(|\mathbf{r}'|-r) \, d^2\mathbf{r}' \tag{3}$$

This is to be accessed via its Fourier transform, specifically:

$$\overline{P}(\mathbf{r},t) = \frac{1}{2\pi D} \int_{-\infty}^{\infty} \Psi(\mathbf{r},\omega)\exp(-i\omega t)d\omega = \frac{1}{2\pi D}\int_{\mathsf{V}} \Psi(\mathbf{r},-ik^2 D)\exp(-k^2 Dt)(-2iD)kdk$$

$$= \frac{1}{i\pi}\int_{-\infty}^{\infty} \Psi(\mathbf{r},-ik^2 D) \, \exp(-k^2 Dt) \, kdk \tag{4}$$

for $t>0$, or zero for $t<0$. Here $\mathsf{V}$ denotes a symmetrically bent contour of integration arising from the change of variables $i\omega = Dk^2$. It runs from left to right, with a right angle bend at the origin. It can be straightened to become the real axis for $t>0$, and fully folded up vertically for $t<0$, yielding zero. It is the wave $\Psi(\mathbf{r},-ik^2 D)$ that obeys the wave equation $\nabla^2 \Psi + k^2 \Psi = -\delta(r-R)/(2\pi R)$ mentioned in the introduction (being the Fourier transform of the diffusion equation). It has a continuous ring source at radius $R$, characterized by having a jump of radial gradient of $(-1/2\pi R)$ across the circle of radius $R$. It is to be solved with Neumann boundary conditions (zero gradient) on the disc radius $a$. The circular symmetry means the solution is everywhere a linear combination of the two Bessel functions $J_0$ and $Y_0$. Also required, as explained in the introduction, is that the waves must be outgoing in nature outside radius $R$ ($\sim H_0^{(1)}(kr) = J_0(kr) + iY_0(kr)$) consistent with there being a source of waves and having zero value of the probability density for $t<0$. Inside the source ring, radius $R$, the wave has to be a (different) linear combination of the Bessel functions $J_0(kr)$ and $Y_0(kr)$ satisfying the Neumann boundary condition (the presence of the disc means that there is no need to exclude the Bessel function $Y_0(kr)$ since its infinity at the origin is inside the disc). The result is the rather formidable looking expressions for $\Psi$ (soon to be simplified):

$$\Psi = J_0(kr)\left[\frac{H_0^{(1)}(kR)(\partial/\partial a)Y_0(ka)}{4i(\partial/\partial a)H_0^{(1)}(ka)}\right] - Y_0(kr)\left[\frac{H_0^{(1)}(kR)(\partial/\partial a)J_0(ka)}{4i(\partial/\partial a)H_0^{(1)}(ka)}\right] \quad \text{for } r<R,$$

$$\Psi = H_0^{(1)}(kr) \left[ \frac{J_0(kR)(\partial/\partial a)Y_0(ka) - Y_0(kR)(\partial/\partial a)J_0(ka)}{4i(\partial/\partial a)H_0^{(1)}(ka)} \right] \quad \text{for } r > R. \quad (5)$$

The Wronskian relation $J_0(kR)(\partial/\partial R)Y_0(kR) - Y_0(kR)(\partial/\partial R)J_0(kR) \equiv 2/(\pi R)$ has already been used to simplify the (common) denominator. With its Bessel functions of $kr$, (5) satisfies the wave equation $\nabla^2 \Psi + k^2 \Psi = 0$ other than at $r=R$. It is constructed to have zero normal derivative at $r=a$, and is outgoing at infinity. Also it is straightforward to show, that it has the required jump of gradient $(-1/2\pi R)$ at $r=R$. As a wave it is complex valued, but, as in the disc-free example in appendix B, some algebra shows that the imaginary part is an even function of $k$ so that, multiplied by $\exp(-k^2 Dt)kdk$ and integrated over all $k$, it is eliminated. The real part has odd and even pieces. Only the odd piece contributes, and it can be extracted using $J_0(-z) = J_0(z)$ and $Y_0(-z) = Y_0(z) + 2iJ_0(z)$ for $z>0$. The unit step functions implicit in (5), distinguishing $r<R$ and $r>R$ disappear, and the result is a real integral over positive $k$ only

$$\overline{P}(\mathbf{r},t) \equiv \frac{1}{2\pi} \int_0^\infty \frac{\left(J_0(kr)Y_1(ka) - Y_0(kr)J_1(ka)\right)\left(J_0(kR)Y_1(ka) - Y_0(kR)J_1(ka)\right)}{J_1(ka)^2 + Y_1(ka)^2} \times$$

$$\times \exp(-k^2 Dt)kdk \quad (6)$$

As a brief check on this, for a zero radius cylinder ($a=0$) one has (since $Y_1$ infinitely dominates both numerator and denominator, and cancels) (6) reduces to

$$\overline{P}(\mathbf{r},t) = \frac{1}{2\pi} \int_0^\infty J_0(kr) J_0(kR) \exp(-k^2 Dt)k \, dk = \frac{1}{2\pi} \frac{1}{2Dt} \exp\left(-\frac{(r^2+R^2)}{4Dt}\right) I_0\left(\frac{rR}{2Dt}\right) \quad (7)$$

[DLMF 2017 formula 10.22.67] which is a long known result e.g.[Kleinert 2006] for $\overline{P}$ with no cylinder present, but still with azimuthal averaging. A one-line derivation of it (A1) is supplied as part of the 'truncated free space model', appendix A.

Proceeding to the formula for $\langle \phi^2 \rangle$, only the exponential in (6) depends on time so the time integration from 0 to $t$ in (2) can be implemented straightaway, and only the single $J_0(kr)$ and $Y_0(kr)$ depend on $r$. Therefore, from (2) and (6) we have our exact result (with sample graphs given as the solid curves in fig 1):

$$\langle\phi^2\rangle \equiv 2\int_0^\infty \frac{dk}{k}\left(1-\exp(-k^2 Dt)\right) \times$$

$$\left(\frac{\left\{\int_a^\infty J_0(kr)dr/r\right\}Y_1(ka) - \left\{\int_a^\infty Y_0(kr)dr/r\right\}J_1(ka)}{J_1(ka)^2 + Y_1(ka)^2}\right) \times \left[J_0(kR)Y_1(ka) - Y_0(kR)J_1(ka)\right]$$

(8)

The two braced functions have no name, but are recognized functions (of *ka* here) with documented convergent series [DLMF 2017 formulas 10.22.39 and 10.22.40] and straightforward asymptotic expansions. The three factors in the integrand separated by the symbols × are of the respective forms: a function of *k* and *t*, of *k* and *a* (in fact of *ka*), and of *k* and *a* and *R*. These will play a part in the evaluation of the *k* integral in the limit described in the next section. First though, there follows a brief comment on each of the three factors separated by the two symbols ×.

In front of the first × symbol there is an everywhere positive function of *k* (>0) (as well as of time *t*; this is the only appearance of *t*). Graphed against *k* it has a single extremum and infinite area under it due to the long tail. For small and large *k* respectively it has the forms $kDt$ and $1/k$.

Next the function of the product *ka* (large bracket) is an everywhere negative function. Graphed against *k* it has a single extremum and finite area ($-\frac{1}{2}\pi/a$). For small and large *ka* respectively it has the forms $\frac{1}{2}\pi ka(\gamma + \log(\frac{1}{2}ka))$ and $-2/(ka)^2$ where $\gamma$ is Euler's constant.

Finally, after the second × symbol, the function of the three quantities *k*, *a*, *R*, (the square bracket) has the small *k* form ($-2/\pi ka$) coming from the $Y_1$ Bessel function, and is oscillatory with amplitude decaying proportional to $1/k$ for large *k*. This square bracket factor contains the only *R* dependence of the integrand.

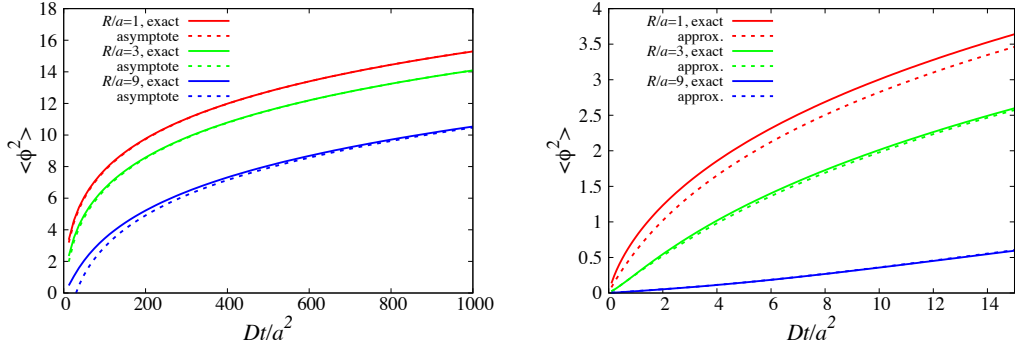

*Fig 1*. Mean square winding angle (radians$^2$) as a function of time (specifically $Dt/a^2$). The two pictures show different time ranges, and for each there are three pairs of curves shown, appropriate to values of the ratio $R/a$ = 1,3,9 of start radius to disc radius. For the left picture each pair of graphs compares the exact formula (8) with the truncated free space model (A2). Not shown, since they coincide visually with the exact curves, are numerical simulations of the Brownian motion. For the right picture each pair of graphs compares the exact formula (8) with the long time asymptotic formula (13). Not shown are curves for the truncated free space model (A2), since, over most of the plot, they coincide visually with the exact curves.

*Limit of narrow cylinder $a^2 << R^2$, highlighting short and long times*

The integrand in (8) for the mean square winding angle simplifies in the limit of narrow cylinder: *a*(cylinder radius)<<*R* (starting radius) allowing analytic (albeit obscure) evaluation. The limits of small and long diffusion times yield results in terms of elementary functions. First the scaled integration variable $\rho=kR$ can be introduced, then the quantity *ka* becomes $\rho a/R$ and the quantity $k\sqrt{Dt}$ becomes $\rho\sqrt{Dt}/R$. The two Bessel functions $J_0(\rho)$ and $Y_0(\rho)$ in the square bracket of (8) have a width scale of order unity, decaying and oscillating for large $\rho$. Thus the functions of *ka*, namely $J_0(ka)$ and $Y_0(ka)$ and the large bracket in (8), are all explored only for small values of their argument $\rho a/R$. Substituting the forms given in the previous section for this small argument limit:

$$\langle \phi^2 \rangle \approx$$
$$2\int_0^\infty \frac{1-\exp(-\rho^2 Dt/R^2)}{\rho} \left(\frac{\pi\rho a}{2R}(\gamma + \log(\frac{\rho a}{2R}))\right) \left[J_0(\rho)\left(-\frac{2R}{\pi\rho a}\right) - Y_0(\rho)\frac{\rho a}{2R}\right] d\rho \qquad (9)$$

Further, the $Y_0$ term can be ignored in the small $a$ limit because of its $a^2$ prefactor (both of the integrals $\int_0^\infty Y_0(\rho)\rho d\rho$ $(= -2/\pi)$ and $\int_0^\infty Y_0(\rho)\rho\log(\rho)d\rho$ are finite numbers if naturally interpreted). In contrast the $J_0$ term tends to infinity from the $\log(a)$ term.

This integral can then be evaluated exactly, though it is rather long and involves one somewhat obscure term at the end:

$$\langle \phi^2 \rangle \approx 2\int_0^\infty \left(\frac{-\gamma - \log(\frac{1}{2}\rho a/R)}{\rho}\right)\left(1 - \exp(-\rho^2 Dt/R^2)\right) J_0(\rho)\, d\rho \qquad (10)$$

$$= 2(-\gamma - \log(a/2R))\, E_1(R^2/4Dt)$$

$$-\frac{1}{4}\left(-\frac{\pi^2}{6} + (\gamma - \log(4))^2 + \log(R^2/Dt)(2\gamma + \log(R^2/16Dt))\right) \qquad (11)$$

$$+ \frac{1}{2}(\gamma - \log(R^2/Dt))\, E_1(R^2/4Dt) + \frac{1}{4}L''(-R^2/4Dt)$$

where the double prime means second derivative of the Laguerre function with respect to order at order equals zero. The bottom two lines of (11) come from the piece of (10) with its first bracket replaced by $-\log(\rho)/\rho$. The top line is simpler and comes from that bracket replaced by $(-\gamma - \log(a/2R))/\rho$. Perhaps most interesting are the short and long diffusion time limits of mean square winding angle.

First the limit of the mean square winding angle around a narrow cylinder (10) for short diffusion time $t$ is evaluated. Part of it, $2(-\gamma - \log(a/2R))\, E_1(R^2/4Dt)$, from the top line of (11), exposes and resolves the apparent immediate infinity of the mean square winding angle around a line (zero radius cylinder) as a clash of limits. The $-\log(a)$ gives infinity, and the $\exp(-constant/t)/t$ (from $E_1$) gives zero. For any finite time their product gets large for small $a$, but if $t$ is small $a$ needs to be extremely small, of order $\exp(-t\, constant\, \exp(constant/t))$.

The remainder, coming from the lower two lines of (11), though not relevant for the clash of limits, is interesting in that it supplies the expected initial linear growth of the mean square winding angle that is present even if the cylinder radius $a$ is zero. It is most easily accessed from the log part of (10) directly

$$2\int_0^\infty \left(\frac{-\log(\rho)}{\rho}\right)\left(1 - \exp(-\rho^2 Dt/R^2)\right) J_0(\rho)\, d\rho$$

$$\approx -(2Dt/R^2)\int_0^\infty \log(\rho) J_0(\rho) \rho d\rho = 2Dt/R^2 \qquad (12)$$

where the exp has been expanded for small $t$, and the resulting integral has been interpreted in a natural way with a temporary (not shown) Gaussian convergence factor $\exp(-\varepsilon\rho^2)$ yielding $\exp(-1/4\varepsilon)\bigl(-Ei(1/4\varepsilon)-2\log(2\varepsilon)\bigr)/4\varepsilon$. This has the limiting value $-1$ as $\varepsilon$ tends to zero.

Now the large $t$ limit of (11) is considered (that is: the narrow cylinder, long time limit $a^2 \ll R^2 \ll Dt$ of the exact formula (8)). The obscure term $L''(-R^2/4Dt)$ has a large $t$ form $\tfrac{1}{2}(R^2/4Dt)^2$ which vanishes in the limit, so it is ignored like other such terms. The rest of the terms have standard large $t$ expansions yielding

$$\langle \phi^2 \rangle \approx \tfrac{1}{4}\log^2\left(\frac{R^2}{4Dt}\right) + \left(\tfrac{1}{2}\gamma + \log\left(\frac{a}{R}\right)\right)\log\left(\frac{R^2}{4Dt}\right) + \gamma\log\left(\frac{a}{R}\right) + \tfrac{1}{4}\gamma^2 + \tfrac{1}{24}\pi^2 \qquad (13)$$

This result agrees with that from the 'truncated free space' model of [Huber and Wilkinson 2019] (after the correction of an additive constant in the model result). It is re-derived in appendix A, leading to (A5) which is identical to (13).

The leading term in (11) is proportional to the square of $\log(t)$ as anticipated in the introduction. A similar leading term was found by [Rudnick and Hu 1987] for the absorbing cylinder case mentioned earlier. A separate, square of $\log(t)$ dependence, was found early on [Spitzer 1958] for the winding angle probability distribution (rather than its mean square value here). It was found to be a Lorentz (or Cauchy) distribution for the long time limit with a zero radius cylinder. That distribution has infinite mean square, so is not informative for the present non-zero radius cylinder (except insofar as the exact result (8) diverges as $a$ tends to zero; from (6), the integral of $(\overline{P}/r^2)2\pi r dr$ diverges due to small $r$)

*Appendix A. Truncated free space model*

Here we present a careful re-derivation of the narrow cylinder, long time limit of the mean square winding angle around the cylinder in the truncated free space approximation Huber and Wilkinson 2019]. The model initially considers the cylinder absent, finding the azimuthally averaged probability density in free space Brownian motion - an azimuthally smeared Gaussian spreading diffusively

$$\overline{P} = \int_0^{2\pi} \frac{1}{4\pi Dt}\exp\left(-\frac{r^2+R^2-2rR\cos\phi}{4Dt}\right)\frac{d\phi}{2\pi} = \frac{1}{2\pi}\frac{1}{2Dt}\exp\left(-\frac{(r^2+R^2)}{4Dt}\right)I_0\left(\frac{rR}{2Dt}\right)$$
(A1)

(This formula appears above, in (7), as the $a=0$ case of the general formula (6)). Now the cylinder is introduced as a virtual presence. Some of the probability density is inside it as well as outside it, which is an unphysicality of the model, though after a long time this inside probability diminishes towards zero. The model simply treats the inside probability as zero, truncating the free space probability density. The mean square winding angle is then, using (2) with (7) for $\overline{P}$ instead of (6):

$$\left\langle \phi^2 \right\rangle = \int_0^t dt' 2D \int_{|\mathbf{r}|>a} \overline{P}_0(\mathbf{r},t'|\mathbf{R}) \frac{d^2\mathbf{r}}{r^2}$$

$$= \int_0^t dt' 2D \int_a^\infty \frac{1}{2\pi} \frac{1}{2Dt'} \exp\left(-\frac{(r^2+R^2)}{4Dt'}\right) I_0\left(\frac{rR}{2Dt'}\right) \frac{2\pi r \, dr}{r^2} \quad (A2)$$

It is this integral form that is used for the graphical comparison with the exact formula (8) in fig 1.

The next step rewrites (A2) with a trick of a standard type, adding and subtracting an easy term (with a view to allowing an integral to be extended to the whole space).

$$\left\langle \phi^2 \right\rangle = \int_0^t \frac{dt'}{t'} \int_a^\infty \exp\left(-\frac{(r^2+R^2)}{4Dt'}\right) \left[I_0\left(\frac{rR}{2Dt'}\right) - 1\right] \frac{dr}{r}$$
$$+ \int_0^t \frac{dt'}{t'} \int_a^\infty \exp\left(-\frac{(r^2+R^2)}{4Dt'}\right) \frac{dr}{r} \quad (A3)$$

The second line here is the main contribution to $\left\langle \phi^2 \right\rangle$ requiring the analysis that follows below. It tends to infinity as $t$ tends to infinity or as $a$ tends to zero (the former being expected for on-going diffusion, and the latter being the striking infinity for a narrow line-like cylinder highlighted in the main text). The first line of (A3), on the other hand, is upper bounded by a constant, $\pi^2/12$, which is the value obtained in the relevant limit where $t=\infty$ and $a=0$. (Then the $t'$ integral, gives $\log\left(2C/(C+\sqrt{C^2-1})\right)$ where $C = \frac{1}{2}(r/R + R/r)$ and the $r$ one yields the $\pi^2/12$ in the next equations (A4) and (A5)).

$$\left\langle \phi^2 \right\rangle \approx \frac{\pi^2}{12} + \int_0^t \frac{dt'}{t'} \int_a^\infty \exp\left(-\frac{(r^2+R^2)}{4Dt'}\right) \frac{dr}{r}$$
$$= \frac{\pi^2}{12} + \int_0^t \frac{dt'}{2t'} \exp\left(-\frac{R^2}{4Dt'}\right) E_1\left(\frac{a^2}{4Dt'}\right) \quad (A4)$$

$$= \frac{\pi^2}{12} + \int_{R^2/4Dt}^{\infty} \frac{\exp(-s)}{2s} E_1\left(\frac{a^2}{R^2}s\right) ds$$

$$\approx \frac{\pi^2}{12} + \int_{R^2/4Dt}^{\infty} \frac{\exp(-s)}{2s} \left(\gamma + \log\frac{a^2}{R^2} + \log(s)\right) ds$$

$$= \frac{\pi^2}{12} + \left(\gamma + \log\frac{a^2}{R^2}\right) E_1\left(\frac{R^2}{4Dt}\right) + \int_{R^2/4Dt}^{\infty} \frac{\exp(-s)}{2s} \log(s) \, ds \qquad (A5)$$

$$= \frac{\pi^2}{12} + \left(\gamma + \log\frac{a^2}{R^2}\right) E_1\left(\frac{R^2}{4Dt}\right)$$

$$\qquad + \left(\left[\tfrac{1}{4}\exp(-s)\log^2(s)\right]_{R^2/4Dt}^{\infty} + \tfrac{1}{4}\int_{R^2/4Dt}^{\infty} \exp(-s)\log^2(s)\,ds\right)$$

Here, after the top line, there is only one further approximation step: that in the fourth line, replacing $E_1$ by the first two terms of its series. The other steps are: introducing a changed variable to get the third line and integrating by parts (differentiating the exp part) to get the fifth line. Finally in the sixth line, we can approximate $E_1$ as before, and take the lower limits in the long bracket as zero with vanishing error as $t$ tends to infinity. The square brackets term then vanishes, and the final integral has the known result $\pi^2/6 + \gamma^2$. In all one has

$$\langle \phi^2 \rangle \approx \tfrac{1}{4}\log^2\left(\frac{R^2}{4Dt}\right) + \tfrac{1}{2}\gamma \log\left(\frac{R^2}{4Dt}\right) + \log\left(\frac{a}{R}\right)\log\left(\frac{R^2}{4Dt}\right) + \gamma \log\left(\frac{a}{R}\right) + \frac{\gamma^2}{4} + \frac{\pi^2}{24}$$
(A6)

This reproduces the result eqn (23) of [Huber and Wilkinson 2019] apart from the additive constant $\pi^2/12$ (from (A4)) which was missed in that analysis. With the corrected constant, the result (A6) for the narrow cylinder, long time, limit of the truncated free space model (A2), equals that, (13), from the exact formula (8) in the same limit.

*Appendix B; Free space with initial ring*

This example is of an initial probability density that is a sharp circular ring or ridge at radius $R$ from the origin $\delta(r-R)/(2\pi R)$ diffusing unobstructed in the plane. The example contains, in a less complicated setting, all the manoeuvres are relevant for the main text problem of diffusion obstructed by the impenetrable disc (projected cylinder). It makes appearance elsewhere, once as the zero radius disc limit (8) of the general formula (7) for probability density, and again as a starting point, (A1), in the analysis in the 'truncated free space' of appendix A. In the latter case it is generated by superposing an infinity of point sources, as in (A1), around a circle. But to represent it again in terms of waves (which we need in the obstructed problem), one seeks the time Fourier transform $\Psi(\mathbf{r},\omega) = \int_{-\infty}^{\infty} \overline{P}(\mathbf{r},t)\exp(i\omega t)dt$, of the spreading probability density (now inwards as well as outwards from the initial ring).

This wave obeys the Helmholtz wave equation with the steady ring of sources on the right hand side $\nabla^2\Psi + k^2\Psi = -(1/2\pi R)\,\delta(r-R)$. The solution $\Psi$ is now a combination of Bessel and Hankel functions of order zero with the following properties. The ring delta function on the right, coming from the analogous term in the diffusion equation, means that $\Psi$, though continuous everywhere, has a discontinuity of gradient of $(-1/2\pi R)$ across the ring at $r=R$. As before, to correspond to there being a source, it must be outgoing at infinity, indeed everywhere outside the ring, thus $\sim H_0^{(1)}(kr)$. Inside the ring it must be smooth at the origin, thus $\sim J_0(kr)$, without any $Y_0(kr)$ which is infinite at the origin. This part of the wave in the inner region $r<R$, can be interpreted as an equal superposition of incoming and outgoing waves. The solution with these features is

$$\Psi = \frac{i}{4}\left(J_0(kr)H_0^{(1)}(kR)\Theta[R-r] + H_0^{(1)}(kr)J_0(kR)\Theta[r-R]\right) \tag{B1}$$

with $\Theta$ as the unit (Heaviside) step function. Its gradient change at the ring is $(ik/4)\left(J_1(kR)H_0^{(1)}(kR) - H_1^{(1)}(kR)J_0(kR)\right)$, in which the large bracket is a Wronskian $(2i/\pi kR)$, so this correctly equals $(-1/2\pi R)$. The consequent probability density is obtained, analogously to (4), by multiplying by the Gaussian and integrating (again the ∨ bending of the contour supplies the time Fourier transform of the probability density).

$$\overline{P}(\mathbf{r},t) = \frac{1}{i\pi}\int_{-\infty}^{\infty}\frac{i}{4}\left(J_0(kr)H_0^{(1)}(kR)\Theta[R-r] + H_0^{(1)}(kr)J_0(kR)\Theta[R-r]\right)\exp(-k^2 Dt)k\,dk \tag{B2}$$

The long bracket simplifies because its part even in $k$ cancels in the integration. Using $H_0^{(1)}(\zeta) = J_0(\zeta) + iY_0(\zeta)$ and $J_0(-\zeta) = J_0(\zeta)$, and $Y_0(-\zeta) = Y_0(\zeta) + 2iJ_0(\zeta)$ (for $\zeta>0$ and real), the unit step functions disappear (since $\Theta[R-r] + \Theta[r-R] = 1$) and the integral reduces to one over positive $k$ only

$$\frac{1}{2\pi}\int_0^{\infty} J_0(kr)\,J_0(kR)\,\exp(-k^2 Dt)k\,dk \tag{B3}$$

(the evaluation of which is given in (6)). That completes the obstruction-free preliminary example.

*Appendix C: Annulus alternative for the diffused ring probability density $\overline{P}(\mathbf{r},t)$.*

An alternative approach to deriving $\overline{P}(\mathbf{r},t)$ that avoids the complex wave scattering is noted here. Instead it involves taking the limit of a finite domain: an annulus with an indefinitely large outer circular boundary of radius $L$, as well as the inner disc one of radius $a$. The diffusion equation then has discrete eigenvalues, and the limit $L \to \infty$, when they are dense, is needed. The probability density is given exactly in terms of the (orthonormal) eigenfunctions $\psi_n(r) = \alpha_n J_0(k_n r) + \beta_n Y_0(k_n r)$, and eigenvalues $k_n$ (with decay eigenfrequencies $Dk_n^2$):

$$\overline{P}_{annulus}(\mathbf{r},t) = \sum_n p_n [\alpha_n J_0(k_n r) + \beta_n Y_0(k_n r)] \exp(-k_n^2 Dt) \tag{C1}$$

with the real coefficients $\alpha_n$ and $\beta_n$ and $p_n$ and $k_n$ to be found. The Neumann boundary conditions at $r=a$ and $r=L$ are $\alpha_n J_1(k_n a) + \beta_n Y_1(k_n a) = 0$ and $\alpha_n J_1(k_n L) + \beta_n Y_1(k_n L) = 0$. These determine the eigenvalues $k_n$ as follows.

Dividing each of these two boundary condition equations by $\sqrt{\alpha_n^2 + \beta_n^2}$ one has the simultaneous equations $\cos\gamma\, J_1(kL) + \sin\gamma\, Y_1(kL) = 0$ and $\cos\gamma\, J_1(ka) + \sin\gamma\, Y_1(ka) = 0$, where a 'phase angle' $\gamma$ has been introduced. The solutions are pairs of numbers $k_n, \gamma_n$, represented graphically as the intersections of the zero contours in the $k, \gamma$ plane of each of the two functions on the left hand sides of the simultaneous equations. Fig 3 shows, for example, the eigenvalues $k_n$ for an annulus with $a=1$ and $L=5$. The intersections come in degenerate (vertically separated) pairs corresponding to ± the same eigenfunction. The lowest eigenvalue, call it $k_0$, is zero, and its eigenfunction is a constant $1/\sqrt{\pi(L^2 - a^2)}$.

Higher eigenvalues $k_n$ with $n \geq 1$ (which have eigenfunctions with $n$ concentric nodal circles in the annulus), become more and more evenly spaced in $k$ as the contours become straighter forming a uniform lattice. The straightness and spacing come from the trigonometric form of $J_0$ and $Y_0$ for large argument $J_1(x) \approx \sqrt{2/\pi x} \cos(x - 3\pi/4)$, and $Y_1(x) \approx \sqrt{2/\pi x} \sin(x - 3\pi/4)$. The simultaneous equations thus become $\cos(kL - \gamma - 3\pi/4) = 0$ and $\cos(ka - \gamma - 3\pi/4) = 0$; then the gradients of the two types of contour $d\gamma/dk$ are respectively equal to $L$ and $a$. The ultimate spacing $\Delta k$ is given by setting the difference of these gradients times $\Delta k$ equal to $\pi$, thus $\Delta k = \pi/(L-a)$.

The normalization condition for each eigenstate is

$$\int_a^L [\alpha_n J_0(k_n r) + \beta_n Y_0(k_n r)]^2 2\pi r dr = 1 \tag{C2}$$

For $n=0$ (with its $k_0=0$), $\beta_0=0$ and $\alpha_0 = 1/\sqrt{\pi(L^2 - a^2)}$. For $n=1$ the values $\alpha_1$ and $\beta_1$ are complicated since $k_1 L$ is of order unity, not large or small as $L \to \infty$. The same holds for all low values of $n$, indeed any fixed value of $n$. However as $L \to \infty$ the eigenvalues $k_n$ become dense (the steep set of lines in fig 2 becoming steeper and more closely packed). In the absence of bad behaviour of the normalization for the low states, any finite number of them can be ignored. For any fixed $k$ (not $n$), in the large $L$ limit the integrals (C2) are (infinitely) dominated by their slowly decaying large $r$ tails. One has for the three terms in the integrand of (C2): $\int_a^L J_0^2(k_n r)2\pi r dr \approx \int_a^L Y_0^2(k_n r)2\pi r dr \approx 2L/k_n$ and the cross term is of order unity and therefore negligible. Combining this with the Neumann condition at r=a, one has for large n

$$\alpha_n = Y_1(k_n a)\sqrt{k_n/2L}/\sqrt{J_1^2(k_n a)+Y_1^2(k_n a)}$$
$$\beta_n = -J_1(k_n a)\sqrt{k_n/2L}/\sqrt{J_1^2(k_n a)+Y_1^2(k_n a)}$$
(C3)

The Neumann condition at r=L in principle determines the exact values $k_n$ as illustrated in fig 2, but the only feature that will be needed when they become dense as $L\to\infty$ is the density. As has already been noted the asymptotic density (for ka>>1) is $1/\Delta k=(L-a)/\pi$. The density for smaller k values, ka<<1, is $1/\Delta k=L/\pi$ (since the dashed contours are in their curved zone, nearly horizontal). As $L\to\infty$ both extremes are $L/\pi$ to leading order.

Finally to find the $p_n$ one uses the initial condition, a delta function ring at radius r=R. The orthonormality of the eigenfunctions $\psi_n$ means that the delta function can be represented by $\sum_m \psi_m(r')\psi_m(r) = \delta(r-r')/2\pi r'$ with r'=R. Then the coefficients $p_n$ are given by

$$p_n = \sum_m \psi_m(R)\int_a^L \psi_m(r)\psi_n(r)2\pi r dr \qquad (C4)$$

The integral yields $\delta_{mn}$ so $p_n = \psi_n(R) = \alpha_n J_0(k_n R)+\beta_n Y_0(k_n R)$ and thus

$$\overline{P}_{annulus}(\mathbf{r},t) = \sum_n [\alpha_n J_0(k_n R)+\beta_n Y_0(k_n R)][\alpha_n J_0(k_n r)+\beta_n Y_0(k_n r)]\exp(-k_n^2 Dt) \quad (C5)$$

Replacing the sum with an integral $\sum \to \int L/\pi\, dk$ and substituting for $\alpha_n$ and $\beta_n$

$$\overline{P}_{annulus}(\mathbf{r},t) =$$
$$\frac{1}{2\pi}\int_0^\infty \frac{[J_0(kr)Y_1(ka)-Y_0(kr)J_1(ka)][J_0(kR)Y_1(ka)-Y_0(kR)J_1(ka)]}{J_1^2(ka)+Y_1^2(ka)}\exp(-k^2 Dt)k\, dk$$
(C6)

The L has cancelled out and this reproduces (6) for $\overline{P}(\mathbf{r},t)$.

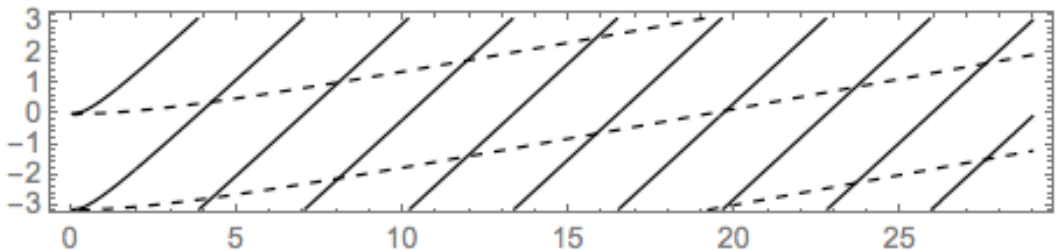

*Fig 2*. Annulus eigenvalues (*a*=1, *L*=5) as the horizontal coordinate *k* of intersections of contours representing boundary conditions at *r=L* and *r=a* (dashed). The vertical coordinate *γ* (with –π<*γ*<π) is an abstract 'phase' as described in the text.

*Acknowledgement*
We are grateful to Baruch Meerson for bringing some of the cited references to our notice.